\documentclass[prd,twocolumn,showpacs,preprintnumbers,amsmath,amssymb,superscriptaddress,floatfix,nofootinbib]{revtex4-2}

\usepackage{graphicx}
\usepackage{bm}
\usepackage{amsmath}
\usepackage{amsfonts}
\usepackage{amssymb}
\usepackage{color}
\usepackage{multirow}
\usepackage{subfigure}
\usepackage{txfonts}
\usepackage[colorlinks, citecolor=blue,anchorcolor=red,menucolor=red, linkcolor=red,filecolor=red,runcolor=red,urlcolor=blue,frenchlinks=red]{hyperref}

\begin{document}

\title{Revisiting the $\Lambda_c^+ \to n\pi^+\eta$ decay in light of the BESIII measurement}

\date{\today}
\author{Meng-Yuan Li}
\affiliation{School of Physics, Beihang University, Beijing 102206, China}
\affiliation{School of Physics, Zhengzhou
	University, Zhengzhou 450001, People’s Republic of China}

\author{Jing Tang}
\affiliation{School of Physics, Zhengzhou
	University, Zhengzhou 450001, People’s Republic of China}

\author{Wen-Tao Lyu}\email{lvwentao9712@163.com}
\affiliation{School of Physics, Zhengzhou
	University, Zhengzhou 450001, People’s Republic of China}
\affiliation{Departamento de Física Teórica and IFIC, Centro Mixto Universidad de Valencia-CSIC Institutos de Investigación de Paterna, 46071 Valencia, Spain}

\author{Shi-Chen Xue}\email{scxue@haut.edu.cn}
\affiliation{School of Physics and Advanced Energy, Henan University of Technology, Zhengzhou, Zhengzhou 450001, People’s Republic of China}

\author{En Wang}\email{wangen@zzu.edu.cn}
\affiliation{School of Physics, Zhengzhou University, Zhengzhou 450001, People’s Republic of China}

\begin{abstract}

Motivated by the recent BESIII measurement, we revisit the $\Lambda_c^+\to n\pi^+\eta$ decay within the chiral unitary approach, including internal and external  $W$-emission mechanisms and final-state interactions. The $a_0(980)^+$ and $N(1535)^{0,+}$ resonances are dynamically generated by meson--meson and meson--baryon interactions, respectively. With the relative weight of internal to external emission fixed at $1/3$, $2/3$, and $1$, all three scenarios reasonably describe the measured $\pi^+\eta$ invariant mass distribution, but yield different fractions for the $a_0(980)$ contribution, and  the fraction obtained for $\gamma=2/3$ and $1$ are numerically below the experimental upper limit. The $a_0(980)$ contribution exhibits a threshold cusp, modified by coherent interference in the total spectrum. Its modest prominence, together with the limited statistical precision and coarse binning, offers a plausible explanation for the absence of a clear experimental signal. We also predict the $\eta n$ and $\pi^+n$ invariant mass distributions. Future high-statistics measurements at BESIII, Belle~II, and STCF would enable a joint analysis to constrain the relative production weight and resonance contributions, providing further insight into the nature of the $a_0(980)$ and $N(1535)$.

\end{abstract}

\maketitle

%%%%%%%%%%%%%%%%%%%%%%
\section{Introduction} \label{sec:Introduction}

The ground-state pseudoscalar and vector meson nonets are well described within the conventional quark model. Nevertheless, the light scalar mesons with quantum numbers $J^{PC}=0^{++}$ exhibit anomalous mass ordering and complex decay behaviors, and their internal structures remain controversial ~\cite{Close:2002zu,Amsler:2004ps,Klempt:2007cp}.
These light scalar mesons, such as $a_0(980)$, $f_0(500)$, $f_0(980)$, $f_0(1370)$, $f_0(1500)$, $f_0(1710)$, and $a_0(1710)$, provide important probes for understanding the nonperturbative nature of Quantum Chromodynamics (QCD), since their mass spectrum and strong couplings to adjacent meson-meson pair thresholds cannot be naturally explained within the conventional $q\bar q$ picture, suggesting possible nonconventional hadronic configurations. 
Various interpretations have been proposed for their internal structures, including multiquark states~\cite{Jaffe:1976ig,Achasov:1999wv,Giacosa:2006rg}, glueball candidates~\cite{Morningstar:1999rf,Chen:2005mg,Cheng:2006zp,Klempt:2007cp}, dynamically generated resonances arising from chiral dynamics~\cite{Janssen:1994wn,Kaiser:1995eg,Oller:1997pn,Pelaez:2021dak,Duan:2020vye,Feng:2020jvp,Wang:2020pem}, and mixtures of different components~\cite{Black:1999yz,Amsler:2004ps}.

Among these light scalar mesons, the $a_0(980)$ is of particular interest because of its proximity to the $K\bar K$ threshold and the interplay between the $\pi\eta$ and $K\bar K$ channels, which make it particularly sensitive to coupled-channel dynamics. Multibody decays of charmed hadrons provide an important platform for investigating these dynamics, as final-state interactions can generate resonant structures in the two-body invariant mass distributions. Recent measurements by Belle, BESIII, and LHCb~\cite{LHCb:2019tdw,Belle:2020xku,BESIII:2023htx,BESIII:2024tpv,BESIII:2024mbf,BESIII:2025yag} have stimulated theoretical studies of $a_0(980)$ production in these decays~\cite{Ikeno:2024fjr,Duan:2024czu,Lyu:2026ack,Lyu:2025oow,Zhang:2025lur,Wang:2022nac,Lyu:2024qgc,Xie:2016evi,Lyu:2024qgc}, offering further opportunities to examine its properties and the role of meson--meson rescattering.

In 2024, within the framework of $\mathrm{SU(3)}$ flavor symmetry, Ref.~\cite{Geng:2024sgq} predicted the branching fraction of the singly Cabibbo-suppressed decay $\Lambda_c^+\to n\pi^+\eta$ to be $(4.52\pm1.21)\times10^{-3}$ without explicitly including intermediate resonances. In our subsequent study within the chiral unitary approach~\cite{Li:2025gvo}, we considered $S$-wave meson--meson and meson--baryon interactions that dynamically generate the $a_0(980)$ and $N(1535)$ resonances, respectively, and obtained a branching-fraction ratio of approximately $0.313$ for the $a_0(980)$-mediated contribution relative to the total decay.

Recently, using a data sample corresponding to an integrated luminosity of $6.1~\mathrm{fb}^{-1}$, the BESIII Collaboration observed the decay $\Lambda_c^+\to n\pi^+\eta$ for the first time and measured its branching fraction to be $(2.94\pm0.59_{\rm stat}\pm0.23_{\rm syst}\pm0.13_{\rm ref})\times10^{-3}$~\cite{BESIII:2026mpt}. This measurement provides a new constraint on the decay dynamics and motivates further investigations of resonance contributions and their interference. Although no significant signal of the intermediate process $\Lambda_c^+\to na_0(980)^+$ was observed in the $\pi^+\eta$ invariant mass distribution, the BESIII Collaboration reported an upper limit of $\mathcal{B}(\Lambda_c^+\to na_0(980)^+)\times\mathcal{B}(a_0(980)^+\to\pi^+\eta)/\mathcal{B}(\Lambda_c^+\to n\pi^+\eta)<0.29$ at the $90\%$ confidence level. As noted by the BESIII Collaboration, this upper limit is close to our earlier theoretical prediction~\cite{Li:2025gvo}.

Despite this experimental progress, the available event sample remains limited, and the measured $\pi^+\eta$ invariant mass distribution is presented with a bin width of approximately $33~\mathrm{MeV}$~\cite{BESIII:2026mpt}. The limited statistical precision and relatively coarse binning hinder the identification of a possible $a_0(980)$ threshold structure and the extraction of its contribution in the presence of interference. Moreover, experimental information on the $\eta n$ and $\pi^+n$ invariant mass distributions is not yet available, leaving the relative strengths of the different production and rescattering mechanisms insufficiently constrained. Nevertheless, the existing $\pi^+\eta$ distribution provides an opportunity to examine these mechanisms and their implications for future measurements.

In this work, we revisit the $\Lambda_c^+\to n\pi^+\eta$ decay within the chiral unitary approach, incorporating internal and external $W$-emission mechanisms and final-state interactions in all three two-body subsystems. The $a_0(980)$ and $N(1535)^{0,+}$ resonances are dynamically generated by $S$-wave pseudoscalar meson--meson and pseudoscalar meson--octet baryon interactions, respectively. We consider different values of the relative weight of internal to external emission and determine the overall normalization for each scenario by fitting the $\pi^+\eta$ invariant mass distribution measured by the BESIII Collaboration. We investigate how this relative weight and coherent interference affect the observable $a_0(980)$ structure and the associated fit fraction. Using the fitted parameters, we also predict the $\eta n$ and $\pi^+n$ invariant mass distributions, providing a basis for future joint analyses to further constrain the decay dynamics.

The remainder of this paper is organized as follows. In Sec.~\ref{sec:Formalism}, we introduce the theoretical framework for the $\Lambda_c^+\to n\pi^+\eta$ decay. In Sec.~\ref{sec:Results}, we present and discuss our numerical results. A concise summary is given in the final section.

%%%%%%%%%%%%%%%%%%%%%%
\section{Formalism} \label{sec:Formalism}

\begin{figure*}[htbp]
	% 第一行 4 张图
	\subfigure[]{\includegraphics[scale=0.31]{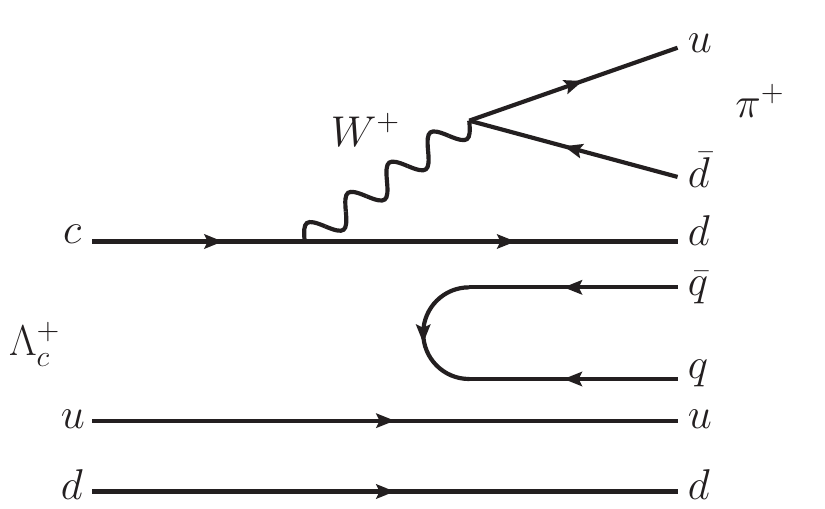}\label{fig:ex1}}
	\subfigure[]{\includegraphics[scale=0.31]{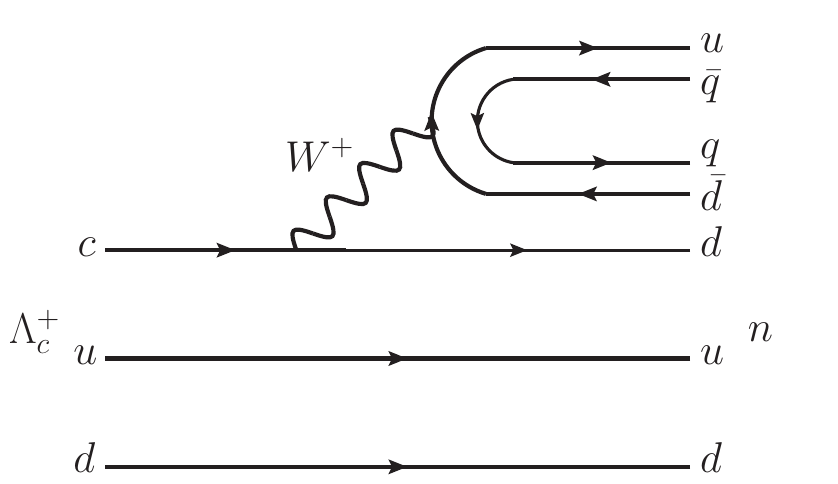}\label{fig:ex2}}
	\subfigure[]{\includegraphics[scale=0.31]{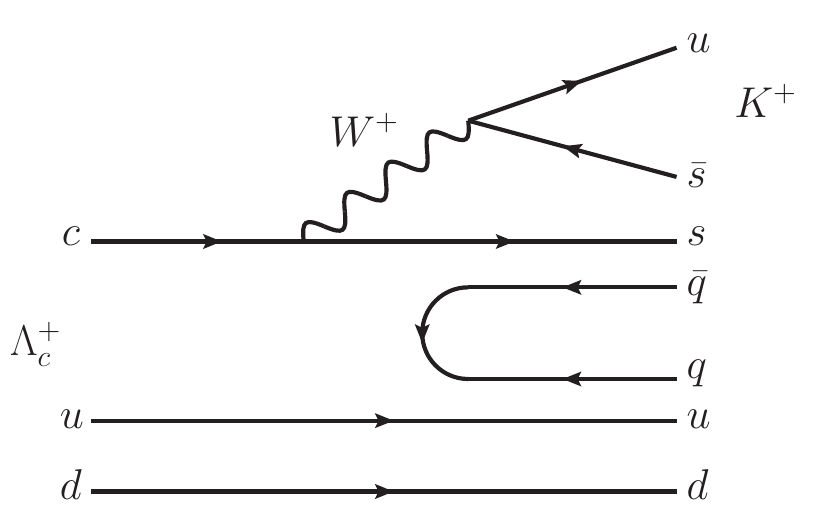}\label{fig:ex3}}
	\subfigure[]{\includegraphics[scale=0.31]{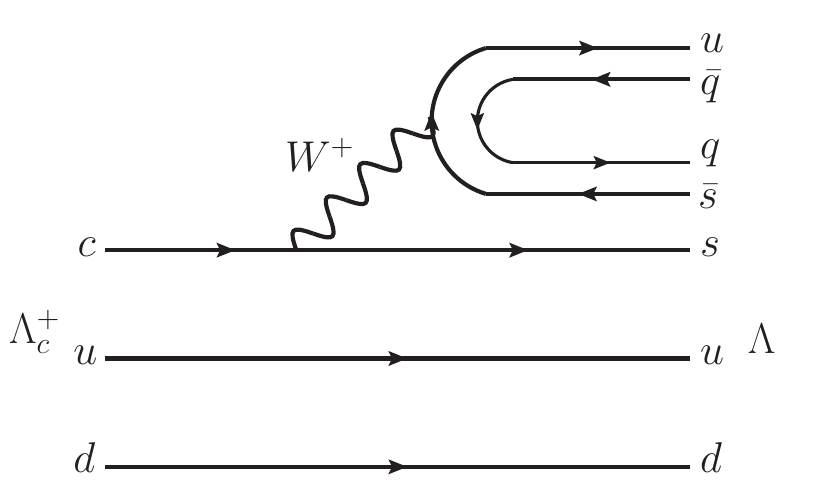}\label{fig:ex4}}
	% 换行
	\vspace{0.5em}
	% 第二行 4 张图
	\subfigure[]{\includegraphics[scale=0.32]{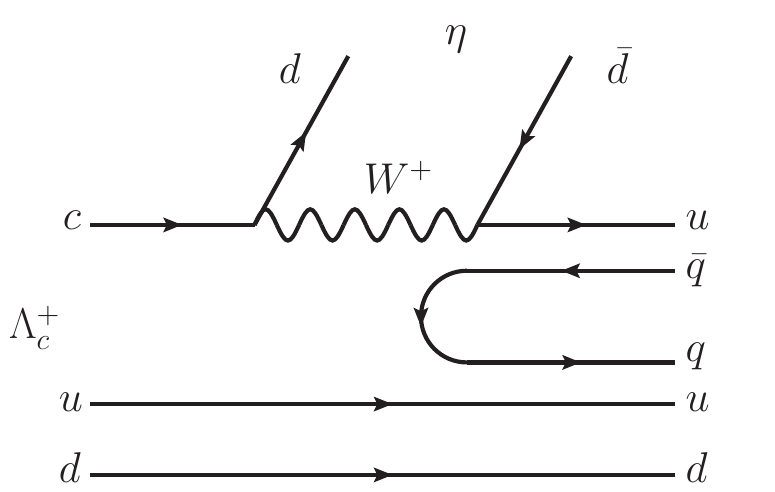}\label{fig:in1}}
	\subfigure[]{\includegraphics[scale=0.32]{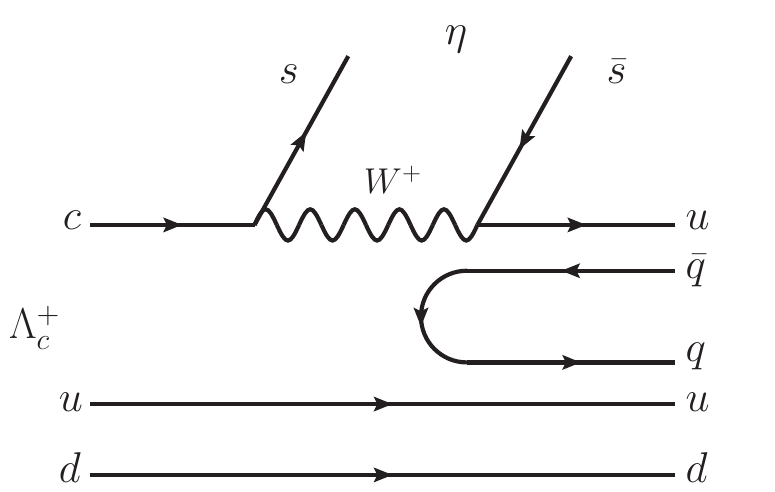}\label{fig:in2}}
	\subfigure[]{\includegraphics[scale=0.32]{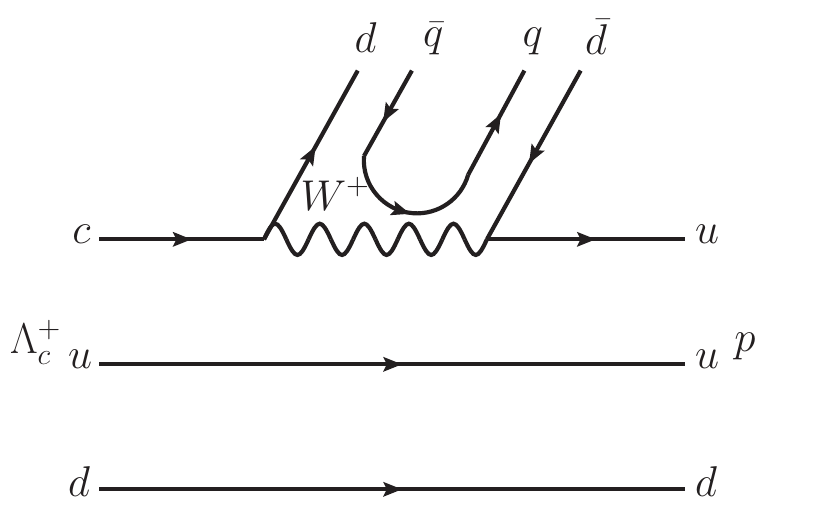}\label{fig:in3}}
	\subfigure[]{\includegraphics[scale=0.32]{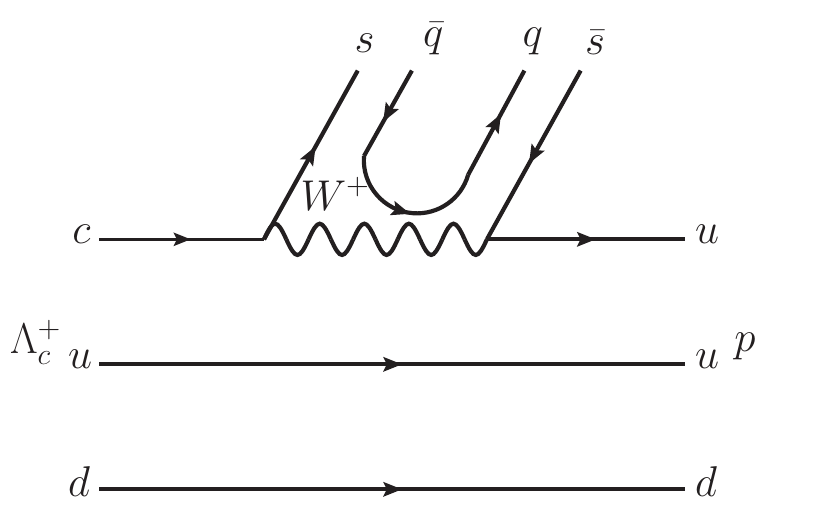}\label{fig:in4}}
	
	\caption{Quark-level diagrams for the processes 
		(a) $\Lambda_c^+ \to \pi^{+} d(\bar{q}q) ud$; 
		(b) $\Lambda_c^+ \to  u(\bar{q}q)\bar{d}~n$;
		(c) $\Lambda_c^+ \to K^{+} s(\bar{q}q) ud$;
		(d) $\Lambda_c^+ \to u(\bar{q}q)\bar{s}~\Lambda$; 
		(e) $\Lambda_c^+ \to (d\bar{d}) u(\bar{q}q) ud$; 
		(f) $\Lambda_c^+\to (s\bar{s}) u(\bar{q}q) ud$; 
		(g) $\Lambda_c^+ \to d(\bar{q}q)\bar{d}~p$; 
		(h) $\Lambda_c^+ \to s(\bar{q}q)\bar{s}~p$. Panels (a)--(d) correspond to the $W^+$ external emission mechanism, while panels (e)--(h) correspond to the $W^+$ internal emission mechanism.}
	\label{fig:external and internal}
\end{figure*}

In this section, we present the theoretical framework for the $\Lambda_c^+ \to  n\pi^+\eta$ decay. The quark-level production mechanisms considered in this work are shown in Fig.~\ref{fig:external and internal}. 
To obtain the corresponding hadronic configurations, we introduce a quark--antiquark pair $\bar q q=\bar u u+\bar d d+\bar{s} s$ with vacuum quantum numbers $J^{PC}=0^{++}$ and carry out hadronization within the SU(3) flavor symmetry framework using the baryon flavor-spin wave functions~\cite{Miyahara:2016yyh,Duan:2024czu,Zhang:2026igc,Lyu:2026ack}.
 
First, we consider the external $W^+$-emission topologies shown in Figs.~\ref{fig:external and internal}\subref{fig:ex1}--\subref{fig:ex4}. For the topology in Fig.~\ref{fig:ex1}, the hadronization yields
\begin{align}\label{eq:ex1}
	H^{(a)}= V_{cd}V_{ud} \frac{\pi^+}{\sqrt{2}} \left( \pi^- p - \frac{1}{\sqrt{2}}\pi^0 n + \frac{1}{\sqrt{3}} \eta n - \frac{2}{\sqrt{6}}K^0 \Lambda \right),  
\end{align}
where \(V_{cd}\) and \(V_{ud}\) are the Cabibbo--Kobayashi--Maskawa (CKM) matrix elements associated with the weak vertices.

For the mechanism in Fig.~\ref{fig:ex2}, the corresponding flavor structure is
\begin{align}       
	H^{(b)} &= V_{cd}V_{ud} \frac{n}{\sqrt{2}} \Bigg\{ \left(  \frac{\eta}{\sqrt{3}} + \frac{\pi^0}{\sqrt{2}}  \right) \pi^+
	+ \pi^+ \left( \frac{\eta}{\sqrt{3}} -\frac{\pi^0}{\sqrt{2}}  \right)
	+ K^+ \bar{K}^0\Bigg\}.
\end{align}
The expression above specifies the flavor structure of the hadronized state. The corresponding weak amplitude also involves the momentum dependence of the $WPP$ vertex. As discussed in Ref.~\cite{Duan:2024okk}, this vertex has the commutator structure $[P,\partial_\mu P]W^\mu$~\cite{Gasser:1983yg,Scherer:2002tk,Ren:2015bsa}, where $P$ denotes the pseudoscalar meson matrix and $W^\mu$ is the weak gauge field. This structure leads to a cancellation between the contributions associated with the $\eta\pi^+$ and $\pi^+\eta$ terms. Following the approximation adopted in Ref.~\cite{Duan:2024okk}, in which the temporal component of the weak current dominates, we also neglect the $K^+\bar K^0$ contribution. Accordingly, this mechanism is not included in the subsequent analysis.

Similarly, hadronization of the mechanisms in Figs.~\ref{fig:ex3} and~\ref{fig:ex4} gives
\begin{align}
	H^{(c)} &=V_{cs}V_{us} \frac{K^+}{\sqrt{2}} \left( K^- p + \bar{K}^0 n + \frac{\sqrt{2}}{3} \eta \Lambda  \right). 
\end{align}
\begin{align}
	H^{(d)} &=V_{cs}V_{us}\left(-\frac{\sqrt{3}}{3}\Lambda \right) \left( \frac{1}{\sqrt{2}}\pi^0 K^+ + \pi^+ K^0\right) .  
\end{align}

The internal $W^+$-emission mechanisms are shown in Figs.~\ref{fig:external and internal}\subref{fig:in1}--\subref{fig:in4}. The corresponding hadronic configurations are
\begin{align}
	H^{(e)} =&V_{cd}V_{ud} \left( \frac{\eta}{\sqrt{6}}  - \frac{\pi^0}{2}  \right) \left( \frac{1}{\sqrt{3}}\eta p  + \frac{1}{\sqrt{2}}\pi^0 p  +  \pi^+ n - \frac{2}{\sqrt{6}} K^+ \Lambda \right), \\
	H^{(f)} =&V_{cs}V_{us} \left( -\frac{\eta }{\sqrt{6}} \right)
	\left( \frac{1}{\sqrt{3}} \eta p +\frac{1}{\sqrt{2}}\pi^0 p  +  \pi^+ n - \frac{2}{\sqrt{6}} K^+ \Lambda \right),  \\
	H^{(g)} =&V_{cd}V_{ud}\frac{p}{\sqrt{2}} \left( \pi^-\pi^+ + \frac{1}{2} \pi^0 \pi^0 + \frac{1}{3} \eta \eta - \frac{2}{\sqrt{6}} \pi^0 \eta + K^0 \bar{K}^0 \right) ,  \\
	H^{(h)} =&V_{cs}V_{us}\frac{p}{\sqrt{2}} \left( K^- K^+ + \bar{K}^0 K^0 + \frac{1}{3} \eta \eta \right) .
\end{align}

Among the hadronic configurations obtained above, we retain those that either directly produce the $n\pi^+\eta$ final state or can reach it through meson--meson or meson--baryon rescattering in a two-body subsystem, with the third hadron acting as a spectator. We introduce an overall normalization constant $V_p$ and a parameter $\gamma$ that characterizes the relative production weight of the internal emission mechanisms with respect to the external emission mechanisms. In general, the eight production topologies in Fig.~\ref{fig:external and internal} may carry different weights. To limit the number of independent production parameters, we assign a common weight to the retained topologies within each emission class and use a single parameter $\gamma$ to specify the relative strength of internal to external emission. This simplified parametrization provides a reasonable description of the available data, while the present statistics do not warrant introducing independent weights for individual topologies. 

Using the approximate CKM relations $V_{cs}\simeq V_{ud}$ and $V_{cd}\simeq -V_{us}$, we absorb the common magnitude of the CKM products into $V_p$ while retaining their relative minus sign. The resulting hadronic configuration is then expressed as   

\begin{align}\label{eq:final state}
	\Lambda_c^+ \Rightarrow& \ 
	V_{p}\Bigg\{-\frac{1}{\sqrt{2}}\pi^+\pi^- p + \frac{1}{2}\pi^+\pi^0 n - \frac{1}{\sqrt{6}}\pi^+\eta n \notag \\
	&+ \frac{1}{\sqrt{2}}K^+\bar{K}^0 n + \frac{1}{3}K^+\eta\Lambda \notag \\
	& +\gamma\Bigg( \frac{1}{2\sqrt{3}}\pi^0\eta p+\frac{1}{2}\pi^+\pi^0n-\frac{\sqrt{2}}{3}\eta\eta p \notag \\
    &-\sqrt{\frac{2}{3}}\pi^+\eta n+\frac{2}{3}\eta K^+\Lambda-\frac{1}{\sqrt{2}}\pi^+\pi^-p \Bigg) \Bigg\}.
\end{align} 

\begin{figure}[htbp]	
	\centering
	\includegraphics[scale=0.4]{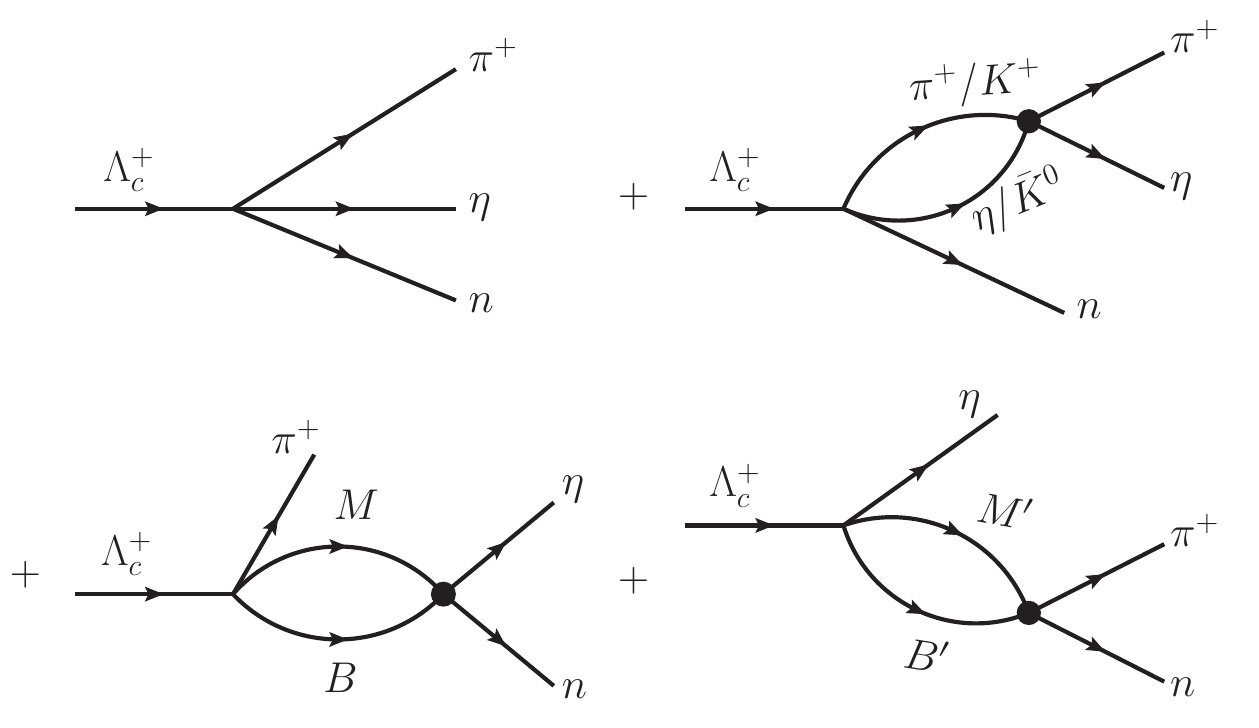}
	\caption{Mechanisms for the tree-level and rescattering processes in the $\Lambda_c^+ \to n\pi^+\eta$ decay.}
	\label{fig:tree_scatter}
\end{figure}

The tree-level and rescattering mechanisms are illustrated in Fig.~\ref{fig:tree_scatter}. The $n\pi^+\eta$ final state can be produced directly at tree-level, with the amplitude
\begin{equation}\label{Eq:M-Tree}
	\mathcal{M}^{\text{Tree}}=V_{p} \Big\{ h_{\pi^{+}\eta n}^{ex}+\gamma h_{\pi^{+}\eta n}^{in} \Big\},
\end{equation}
where $h_{\pi^{+}\eta n}^{\mathrm{ex}}=-1/\sqrt{6}$ and
$h_{\pi^{+}\eta n}^{\mathrm{in}}=-\sqrt{2/3}$ are the production weights given in Eq.~\eqref{eq:final state}. Throughout this work, the superscripts ``ex'' and ``in'' denote external and internal emission, respectively.
 
In addition to the direct production mechanism, the $n\pi^+\eta$ final state can also be reached through meson--meson rescattering. The $a_0(980)$ resonance is dynamically generated by the coupled-channel meson--meson interactions and contributes through the rescattering $\pi^+\eta\to\pi^+\eta$ and $K^+\bar K^0\to\pi^+\eta$. Accordingly, the decay amplitude associated with the $a_0(980)$ resonance can be written as
\begin{align}\label{Eq:M-980}
	\mathcal{M}^{a_0(980)} 
	&= V_{p}\Bigg\{  h_{\pi^{+}\eta n}^{ex} G_{\pi^{+}\eta}(M_{\pi^{+}\eta})t_{\pi^{+}\eta\rightarrow\pi^{+}\eta}  (M_{\pi^{+}\eta})\notag \\
	&+ h_{K^{+}\bar{K}^{0}n}^{ex} G_{K^{+}\bar{K}^{0}}(M_{\pi^{+}\eta})t_{K^{+}\bar{K}^{0}\rightarrow\pi^{+}\eta}(M_{\pi^{+}\eta})\notag \\
    & + \gamma h_{\pi^{+}\eta n}^{in} G_{\pi^{+}\eta}(M_{\pi^{+}\eta})t_{\pi^{+}\eta\rightarrow\pi^{+}\eta}  (M_{\pi^{+}\eta}) \Bigg\},  
\end{align}
where $h_{K^{+}\bar{K}^{0}n}^{ex}=\sqrt{{1}/{2}}$. Here, $G_i$ denotes the meson--meson loop function for $i$ channel, and $t_{i\to j}$ is the $S$-wave transition amplitude from channel $i$ to channel $j$. We consider the two coupled channels $K^+\bar K^0$ and $\pi^+\eta$ and obtain the scattering amplitudes by solving the coupled-channel Bethe--Salpeter equation. Following Refs.~\cite{Xie:2014tma,Li:2025gvo,Zhang:2024myn}, we regularize the meson--meson loop functions using a three-momentum cutoff of $q_{\mathrm{max}}=600~\mathrm{MeV}$.

Meson--baryon final-state interactions also contribute to this decay. With the $\pi^+$ acting as a spectator, the retained $\pi^-p$, $\pi^0n$, and $\eta n$ source channels can rescatter into $\eta n$~\cite{Li:2026lbo}. Likewise, with the $\eta$ acting as a spectator, the $\pi^0p$, $\pi^+n$, $\eta p$, and $K^+\Lambda$ pairs can rescatter into $\pi^+n$. The coupled-channel meson--baryon interactions dynamically generate the $N(1535)^0$ and $N(1535)^+$ resonances, which contribute to the $\eta n$ and $\pi^+n$ final states, respectively. Their corresponding decay amplitudes are given separately as follows
\begin{align}
	\mathcal{M}^{N(1535)^0} = &V_{p}\Bigg\{ \; h_{\pi^{+}\pi^- p}^{ex} \tilde{G}_{\pi^- p}(M_{\eta n}) t_{\pi^- p \to \eta n} (M_{\eta n}) \notag \\
	&+ h_{\pi^{+}\pi^0 n}^{ex}  \tilde{G}_{\pi^0 n} (M_{\eta n}) t_{\pi^0 n \to \eta n}(M_{\eta n}) \notag \\
	&+ h_{\pi^{+}\eta n}^{ex}  \tilde{G}_{\eta n} (M_{\eta n}) t_{\eta n \to \eta n}(M_{\eta n}) \notag \\
	&+ \gamma\Big[ h_{\pi^{+}\pi^0 n}^{in}  \tilde{G}_{\pi^0 n}(M_{\eta n}) t_{\pi^0 n \to \eta n}(M_{\eta n}) \notag\\
    &+ h_{\pi^{+}\eta n}^{in}  \tilde{G}_{\eta n} (M_{\eta n}) t_{\eta n \to \eta n}(M_{\eta n}) \notag \\
	&+  h_{\pi^{+}\pi^- p}^{in} \tilde{G}_{\pi^- p}(M_{\eta n}) t_{\pi^- p \to \eta n}(M_{\eta n})\Big] \Bigg\} , \label{eq:N15350} 
\end{align}
\begin{align}
	\mathcal{M}^{N(1535)^+} = &V_{p}\Big\{ \; h_{\pi^{+}\eta n}^{ex} \tilde{G}_{\pi^+ n}(M_{\pi^+ n}) \, t_{\pi^+ n \to \pi^+ n}(M_{\pi^+ n}) \notag\\
	&+ h_{K^+\eta\Lambda}^{ex} \tilde{G}_{K^+ \Lambda} (M_{\pi^+ n}) \, t_{K^+ \Lambda \to \pi^+ n}(M_{\pi^+ n}) \notag\\
	&+ \gamma\Big[ h_{\pi^{0}\eta p}^{in} \tilde{G}_{\pi^0 p}(M_{\pi^+ n}) \, t_{\pi^0 p \to \pi^+ n}(M_{\pi^+ n})  \notag\\
	& +2 h_{\eta\eta p}^{in} \tilde{G}_{\eta p}(M_{\pi^+ n}) \, t_{\eta p \to \pi^+ n}(M_{\pi^+ n})\notag\\
    & +h_{\pi^{+}\eta n}^{in} \tilde{G}_{\pi^+ n}(M_{\pi^+ n}) \, t_{\pi^+ n \to \pi^+ n}(M_{\pi^+ n}) \notag\\ 
    &+ h_{K^+\eta\Lambda}^{in} \tilde{G}_{K^+ \Lambda} (M_{\pi^+ n}) \, t_{K^+ \Lambda \to \pi^+ n}(M_{\pi^+ n}) \Big] \Bigg\} \label{eq:N1535+} ,
\end{align}
where $h_{i}^{ex/in}$ are the production weights associated with external and internal emission mechanisms, respectively, as specified in Eq.~\eqref{eq:final state}. The factor of $2$ multiplying the $\eta\eta p$ contribution in Eq.~\eqref{eq:N1535+} accounts for the two identical $\eta$ mesons. For the neutral meson--baryon subsystem ($MB$), we include the six channels $K^+\Sigma^-$, $K^0\Sigma^0$, $K^0\Lambda$, $\pi^-p$, $\pi^0n$, and $\eta n$.
For the positively charged subsystem ($M^{\prime}B^{\prime}$), we include $K^0\Sigma^+$, $K^+\Sigma^0$, $K^+\Lambda$, $\pi^0p$, $\pi^+n$, and $\eta p$. Both sets have strangeness $S=0$, and the $N(1535)$ resonances occur in their $I=1/2$ components.

The meson--baryon loop functions appearing in the transition amplitude are evaluated using dimensional regularization, with a renormalization scale $\mu=1200~\mathrm{MeV}$ and subtraction constants $a_{K\Sigma}=-2.8$, $a_{K\Lambda}=1.6$, $a_{\pi N}=2.0$, and $a_{\eta N}=0.2$~\cite{Inoue:2001ip}. The interaction kernel is also taken from Ref.~\cite{Inoue:2001ip} and reads
\begin{align} \label{Eq:vij-pin}
	V_{ij} &= -C_{ij} \frac{1}{4f_i f_j} \bigl(2\sqrt{s} - M_i - M_j\bigr) \notag \\
	&\quad \times \left( \frac{M_i + E_i}{2M_i} \right)^{1/2} \left( \frac{M_j + E_j}{2M_j} \right)^{1/2}.
\end{align} 
Here, $M_i$ and $E_i$ are the baryon mass and energy for channel $i$, respectively, and $f_i$ denotes the corresponding meson decay constant. The coefficients $C_{ij}$ for the neutral and positively charged sectors are taken from Refs.~\cite{Inoue:2001ip,Doring:2005bx}, respectively.

For the meson--baryon loop functions $\tilde G_i$ outside the transition amplitudes in Eqs.~\eqref{eq:N15350} and~\eqref{eq:N1535+}, we employ a three-momentum cutoff regularization scheme with $\tilde q_{\mathrm{max}}=1300~\mathrm{MeV}$. This differs from the dimensional regularization scheme used to calculate the transition amplitudes. Matching the loop functions obtained in these two schemes typically yields negative subtraction constants $a_i(\mu)$ for conventional choices of the cutoff and renormalization scale.
In contrast, positive subtraction constants are adopted for the $K\Lambda$, $\pi N$, and $\eta N$ channels in Ref.~\cite{Inoue:2001ip} to compensate for contributions from reaction channels not explicitly included in the scattering amplitudes. Following the prescription of Refs.~\cite{Wang:2015pcn,Lu:2016roh,Li:2026umb}, we therefore adopt this more natural three-momentum cutoff scheme for these loop functions $\tilde G_i$, which are given by
\begin{align}\label{eq:M-total2}
	\tilde G_i(M_{\mathrm{inv}})
	&=
	\int\frac{d^3q}{(2\pi)^3}
	\frac{M_i}{2\omega_i(q)E_i(q)}
	\frac{1}{
		M_{\mathrm{inv}}-\omega_i(q)-E_i(q)
		+\mathrm{i}\epsilon},
\end{align}
where $E_i(q)=\sqrt{M_i^2+q^2}$ and $\omega_i(q)=\sqrt{m_i^2+q^2}$ are the baryon and meson energies in channel $i$, respectively.

The total decay amplitude is the coherent sum of the tree-level and rescattering amplitudes,
\begin{align}
	\mathcal{M} &=  \mathcal{M}^{\text{Tree}} +  \mathcal{M}^{a_0(980)} + \mathcal{M}^{N(1535)^0}   + \mathcal{M}^{N(1535)^+} .
	\label{eq:amplitude}
\end{align}
The double differential decay width for this decay is given by
\begin{eqnarray}
    \frac{d^{2}\Gamma}{dM_{\eta n}^2 dM_{\pi^{+}\eta}^2}&=\dfrac{1}{(2\pi)^{3}}\dfrac{M_{n}}{8M_{\Lambda_c^+}^2}|\mathcal{M}|^{2}. \label {eq:dgammadm12dm23} 
\end{eqnarray}
By integrating the double differential decay width $\mathrm{d}^2\Gamma/(\mathrm{d}M_{\pi^+\eta}\,\mathrm{d}M_{\eta n})$ over the invariant mass $M_{\eta n}$, we obtain the single differential decay width $\mathrm{d}\Gamma/\mathrm{d}M_{\pi^+\eta}$. The integration interval is constrained by the kinematic boundaries given in the Review of Particle Physics~\cite{ParticleDataGroup:2024cfk}.

%%%%%%%%%%%%%%%%%%%%%%
 \section{Results and Discussion} \label{sec:Results}

\begin{figure*}[htbp]
\subfigure{\includegraphics[scale=0.46]{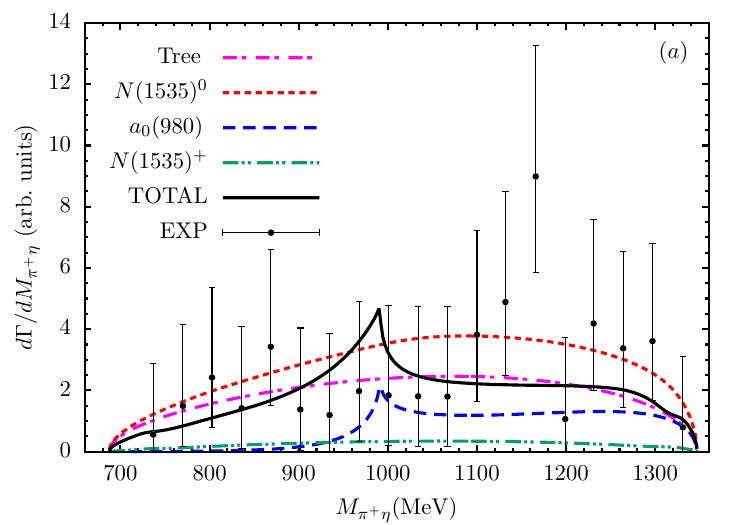}\label{fig:1}}  \subfigure{\includegraphics[scale=0.46]{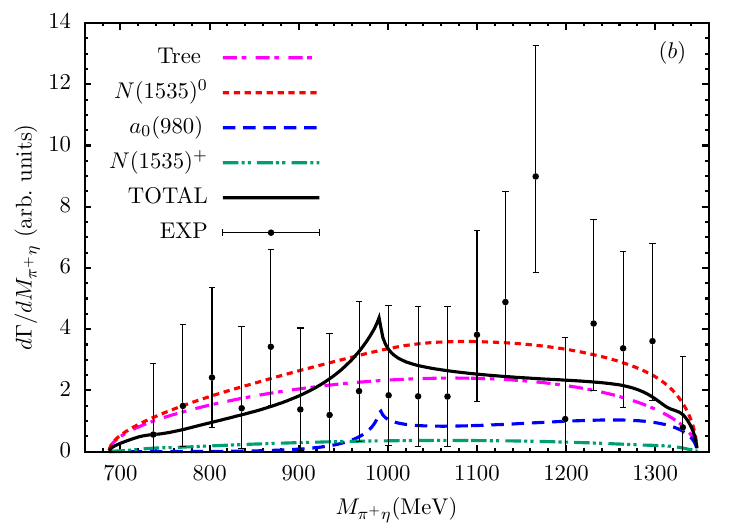}\label{fig:2}} \subfigure{\includegraphics[scale=0.46]{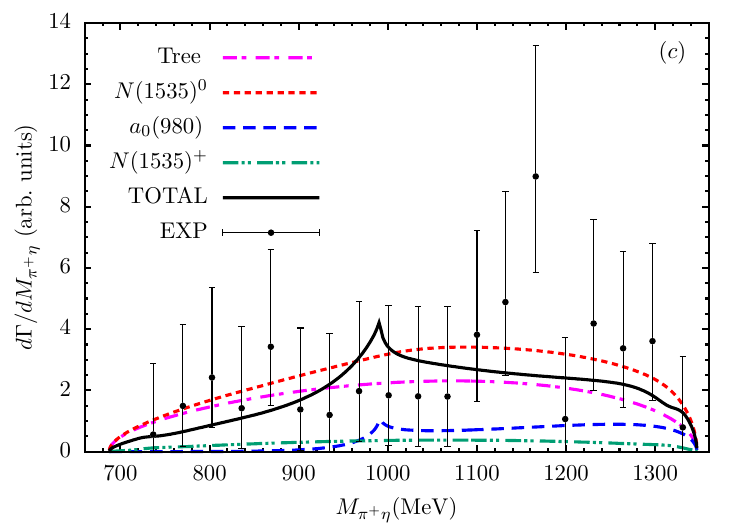}\label{fig:3}}
\caption{$\pi^+\eta$ invariant mass distributions for the  process $\Lambda_c^+ \to n\pi^+\eta$, calculated with fixed different \(\gamma\) values and the corresponding fitted parameters. Panels (a)--(c) correspond to $\gamma=1/3$, $2/3$, and $1$, respectively. The "EXP" denotes the BESIII experimental data~\cite{BESIII:2026mpt}, which are plotted as points with error bars after  background subtraction.}
\label{fig:fit1}
\end{figure*}

 \begin{table}[htb]
\centering
\caption{Fitted parameters for the $\Lambda_c^+ \to n\pi^+\eta$ decay with $\gamma$ fixed at $1/3$, $2/3$, and $1$.}
\begin{ruledtabular}
\begin{tabular}{cccc}
$\gamma$ & $V_p$ & $\chi^{2}/\mathrm{d.o.f.}$  & $R$ \\
\hline
$1/3$ &$0.125\pm0.016$  &0.49 &0.40    \\
$2/3$ &$0.088\pm0.011$  &0.46 &0.28    \\
$1$ &$0.067\pm0.008$  &0.46 &0.24    \\
\end{tabular}
\end{ruledtabular}
\label{tab:fit_gamma}
\end{table}

Naive color counting suggests that internal emission is suppressed relative to external emission by a factor of approximately $1/N_c$, where $N_c=3$ is the number of colors. We therefore take $\gamma=1/3$ as a reference value and also consider $\gamma=2/3$ and $1$ to explore the sensitivity of the results to the relative production weight. The color-counting argument provides a reference scale rather than a quantitative prediction for the hadronic production weight. For each fixed value of $\gamma$, the overall normalization factor $V_p$ is the only free fit parameter. A $\chi^2$ fit to the data of the $\pi^+\eta$ invariant mass distribution from the BESIII Collaboration~\cite{BESIII:2026mpt} yields the fitted normalizations and corresponding $\chi^2/\mathrm{d.o.f.}$ values listed in Table~\ref{tab:fit_gamma}. The resulting distributions are shown in Figs.~\ref{fig:fit1}(a)--(c).

All three scenarios provide reasonable descriptions of the existing data, with comparable fit quality. Figure~\ref{fig:fit1} also shows the individual contributions calculated by retaining each amplitude separately, whereas the total contribution is obtained from their coherent sum. The total distribution therefore includes interference terms and is not simply the sum of the individual component curves.

The individual $a_0(980)$ contribution exhibits a cusp near the $K^+\bar K^0$ threshold, reflecting the coupled-channel dynamics of the $\pi^+\eta$ and $K^+\bar K^0$ systems. This structure remains visible in the total spectrum, although its strength and local shape are modified by interference with the tree-level amplitude and the two $N(1535)$ contributions. The observable manifestation of the $a_0(980)$ thus depends on both its component strength and its interference with the other amplitudes. Its modest prominence, together with the sizable statistical uncertainties and relatively coarse binning of the current measurement, offers a plausible explanation for the absence of a clear experimental signal.

To quantify the $a_0(980)$ contribution within the amplitude decomposition adopted here, we define the fit fraction
\begin{equation}
\begin{aligned}
R&=\dfrac{\displaystyle\int \frac{d^2\Gamma_{a_0(980)}}{dM_{\eta n}\,dM_{\pi^+\eta}}dM_{\eta n}\,dM_{\pi^+\eta}}{\displaystyle\int \frac{d^2\Gamma_{\text{total}}}{dM_{\eta n}\,dM_{\pi^+\eta}}dM_{\eta n}\,dM_{\pi^+\eta}}.
\end{aligned}
\end{equation}
Both integrals extend over the full physical phase space. The numerator is calculated from the $a_0(980)$-associated meson--meson rescattering amplitude alone, while the denominator is obtained from the coherent sum of all contributing amplitudes, including their interference. Thus, $R$ characterizes the strength of this component within our model; it is not a model-independent two-body branching-fraction ratio or a probability for a mutually exclusive decay mechanism.

For $\gamma=1/3$, $2/3$, and $1$, we obtain $R=0.40$, $0.28$, and $0.24$, respectively, as listed in Table~\ref{tab:fit_gamma}. Although the three scenarios yield similar total spectra, their extracted fit fractions differ appreciably. This illustrates the dependence of the component strength on the relative production weight and the resulting interference pattern, and indicates that the current data do not uniquely determine the individual contributions.

The BESIII Collaboration reports the upper limit $\mathcal{B}(\Lambda_c^+\to na_0(980)^+)\times\mathcal{B}(a_0(980)^+\to\pi^+\eta)/\mathcal{B}(\Lambda_c^+\to n\pi^+\eta)<0.29$ at the $90\%$ confidence level~\cite{BESIII:2026mpt}. Numerically, our results for $\gamma=2/3$ and $1$ lie below this limit, whereas that for $\gamma=1/3$ exceeds it. However, the experimental limit was obtained assuming an incoherent sum of the $a_0(980)$ and nonresonant phase-space contributions, while our fit fractions are evaluated within a coherent amplitude model. This comparison therefore provides a reference under different model assumptions and should not, by itself, be interpreted as excluding the $\gamma=1/3$ scenario.

\begin{figure}[tbhp]
	\centering
	\includegraphics[scale=0.6]{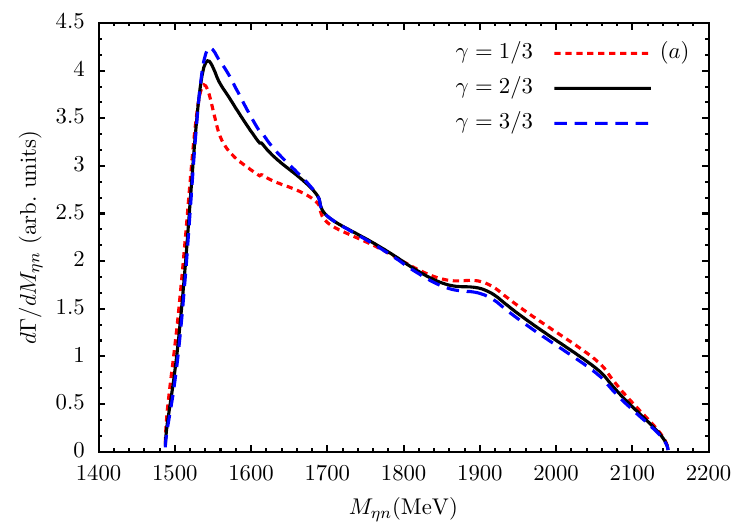}
	\includegraphics[scale=0.6]{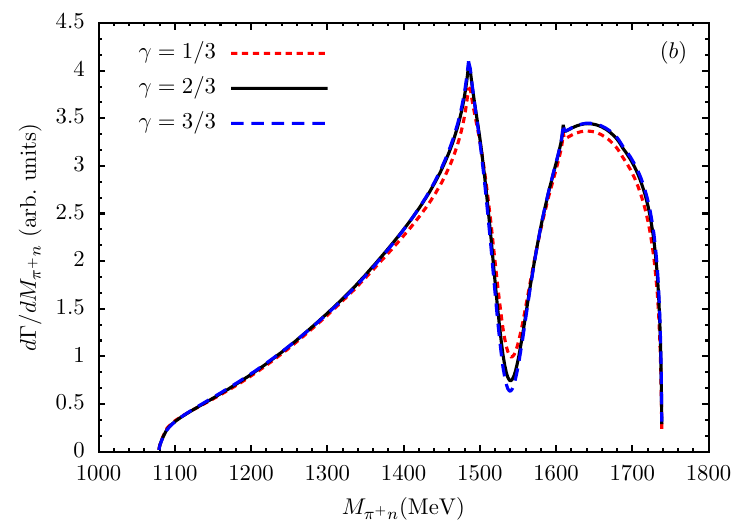}
	\caption{$\eta n$ (a) and $\pi^+n$ (b) invariant mass distributions for the $\Lambda_c^+\to n\pi^+\eta$ decay, calculated using the fit parameters listed in Table~\ref{tab:fit_gamma}.}
	\label{fig:etan_pin}
\end{figure}

Using the fitted normalizations listed in Table~\ref{tab:fit_gamma}, we further predict the $\eta n$ and $\pi^+n$ invariant mass distributions shown in Fig.~\ref{fig:etan_pin}. The $\eta n$ distribution exhibits a pronounced near-threshold enhancement associated with the $N(1535)^0$ contribution, while the $\pi^+n$ spectrum shows a dip around $1535~\mathrm{MeV}$ associated with the $N(1535)^+$ contribution in the coherent total amplitude. 

After refitting $V_p$ for each value of $\gamma$, the predicted $\eta n$ and $\pi^+n$ distributions remain broadly similar. Some local differences occur near the $\eta n$ threshold, whereas the variations in the $\pi^+n$ distribution are smaller. Within the present model and the parameter range examined, the main spectral features are therefore relatively stable against changes in the relative production weight. This weak sensitivity suggests that the overall one-dimensional shapes alone may have limited power to distinguish the three scenarios. Precision measurements of local structures and correlations in the Dalitz distribution would provide complementary tests.

The measured $\pi^+\eta$ invariant mass distribution also shows a tentative local enhancement around $M_{\pi^+\eta}\simeq1160~\mathrm{MeV}$~\cite{BESIII:2026mpt}. Its origin remains unclear given the limited statistical precision and relatively coarse binning. Possible explanations include statistical fluctuations or kinematic reflections of structures in the other two-body subsystems. This feature alone does not provide conclusive evidence for an additional resonance contribution.

Although the mechanisms considered here reasonably describe the available $\pi^+\eta$ spectrum, a single invariant mass projection cannot establish whether additional excited nucleon states are required in the $\eta n$ and $\pi^+n$ channels. The present results should therefore be viewed as predictions within the specified production and rescattering model, rather than as a unique amplitude decomposition of the decay. Future measurements of all three two-body invariant mass distributions and the Dalitz distribution would enable a combined analysis to further constrain the production weights, assess the need for additional relative phases, and investigate possible contributions from other resonances.

%%%%%%%%%%%%%%%%%%%%%%
\section{Summary} \label{sec:Conclusions}
Motivated by the recent BESIII measurement, we have revisited the $\Lambda_c^+\to n\pi^+\eta$ decay within the chiral unitary approach. Our model includes internal and external $W$-emission mechanisms, together with coupled-channel final-state interactions in all three two-body subsystems. The $a_0(980)^+$ and $N(1535)^{0,+}$ resonances are dynamically generated by $S$-wave meson--meson and meson--baryon interactions, respectively.

We have considered three fixed values of the relative production weight, $\gamma=1/3$, $2/3$, and $1$, and determined the overall normalization for each scenario by fitting the $\pi^+\eta$ invariant mass distribution measured by the BESIII Collaboration. All three scenarios provide reasonable descriptions of the existing data, with comparable fit quality. The individual $a_0(980)$ contribution exhibits a cusp near the $K^+\bar K^0$ threshold. This structure remains present in the total spectrum, although its strength and local shape are modified by coherent interference. Its modest prominence, combined with the limited statistical precision and coarse binning of the current measurement, offers a plausible explanation for the absence of a clear experimental signal.

The extracted $a_0(980)$ fit fractions are approximately $0.40$, $0.28$, and $0.24$ for the three scenarios, respectively. The latter two values are numerically below the upper limit of $0.29$ of the BESIII measurement, whereas the first exceeds it. However, this comparison is limited by the different treatments of interference in the experimental extraction and our model. The variation of the fit fractions among scenarios with similar total spectra further demonstrates that the current data do not uniquely constrain the individual component strengths.

Using the fitted normalizations, we have also predicted the $\eta n$ and $\pi^+n$ invariant mass distributions. These exhibit a near-threshold enhancement and a dip around $1535~\mathrm{MeV}$ associated with the $N(1535)^0$ and $N(1535)^+$ contributions, respectively. Their main features remain relatively stable over the range of $\gamma$ considered, providing a basis for future experimental tests of the decay dynamics.

Future high-statistics measurements at BESIII, Belle~II, and STCF, including all three two-body invariant mass distributions and the Dalitz distribution, would enable a combined amplitude analysis to further constrain the relative production weight, possible additional relative phases, and other resonance contributions. Our results illustrate that the absence of a clear $a_0(980)$ signal does not imply the absence of an $a_0(980)$ contribution. Improved measurements and an explicit treatment of interference are essential for determining its role in this decay and testing the underlying coupled-channel dynamics.

\section*{Acknowledgments}
This work was supported by the National Key R\&D Program of China (Grant No. 2024YFE0105200), the National Natural Science Foundation of China (Grant Nos. 12475086, and 12192263), the Natural Science Foundation of Henan Grant No. 252300423951),  and the Zhengzhou University Young Student Basic Research Projects for PhD students (Grant No. ZDBJ202522). Wen-Tao Lyu acknowledges the support of the China Scholarship Council.

%\bibliographystyle{unsrt}
%\bibliography{ref}

\begin{thebibliography}{10}

\bibitem{Close:2002zu}
F.~E.~Close and N.~A.~Tornqvist,
%``Scalar mesons above and below 1-GeV,''
J. Phys. G \textbf{28} (2002), R249-R267
%doi:10.1088/0954-3899/28/10/201
%[arXiv:hep-ph/0204205 [hep-ph]].

\bibitem{Amsler:2004ps}
C.~Amsler and N.~A.~Tornqvist,
%``Mesons beyond the naive quark model,''
Phys. Rept. \textbf{389} (2004), 61-117
%doi:10.1016/j.physrep.2003.09.003

\bibitem{Klempt:2007cp}
E.~Klempt and A.~Zaitsev,
%``Glueballs, Hybrids, Multiquarks. Experimental facts versus QCD inspired concepts,''
Phys. Rept. \textbf{454} (2007), 1-202
%doi:10.1016/j.physrep.2007.07.006
%[arXiv:0708.4016 [hep-ph]].

\bibitem{Jaffe:1976ig}
R.~L.~Jaffe,
%``Multi-Quark Hadrons. 1. The Phenomenology of (2 Quark 2 anti-Quark) Mesons,''
Phys. Rev. D \textbf{15} (1977), 267
%doi:10.1103/PhysRevD.15.267

\bibitem{Achasov:1999wv}
N.~N.~Achasov,
%``On nature of scalar a(0)(980) and f(0)(980) mesons,''
Nucl. Phys. A \textbf{675} (2000), 279C-284C 
%doi:10.1016/S0375-9474(00)00266-9
%[arXiv:hep-ph/9910540 [hep-ph]].

\bibitem{Giacosa:2006rg}
F.~Giacosa,
%``Strong and electromagnetic decays of the light scalar mesons interpreted as tetraquark states,''
Phys. Rev. D \textbf{74} (2006), 014028
%doi:10.1103/PhysRevD.74.014028
%[arXiv:hep-ph/0605191 [hep-ph]].

\bibitem{Morningstar:1999rf}
C.~J.~Morningstar and M.~J.~Peardon,
%``The Glueball spectrum from an anisotropic lattice study,''
Phys. Rev. D \textbf{60} (1999), 034509
%doi:10.1103/PhysRevD.60.034509
%[arXiv:hep-lat/9901004 [hep-lat]].

%\cite{Chen:2005mg}
\bibitem{Chen:2005mg}
Y.~Chen, A.~Alexandru, S.~J.~Dong, T.~Draper, I.~Horvath, F.~X.~Lee, K.~F.~Liu, N.~Mathur, C.~Morningstar and M.~Peardon, \textit{et al.}
%``Glueball spectrum and matrix elements on anisotropic lattices,''
Phys. Rev. D \textbf{73} (2006), 014516
%doi:10.1103/PhysRevD.73.014516
%[arXiv:hep-lat/0510074 [hep-lat]].
%786 citations counted in INSPIRE as of 17 Oct 2026

\bibitem{Cheng:2006zp}
H.~Y.~Cheng,
%``Mixing of Scalar Glueball and Scalar Quarkonia,''
Conf. Proc. C \textbf{060726} (2006), 960-963
%[arXiv:hep-ph/0609229 [hep-ph]].

\bibitem{Janssen:1994wn}
G.~Janssen, B.~C.~Pearce, K.~Holinde and J.~Speth,
%``On the structure of the scalar mesons f0 (975) and a0 (980),''
Phys. Rev. D \textbf{52} (1995), 2690-2700
%doi:10.1103/PhysRevD.52.2690
%[arXiv:nucl-th/9411021 [nucl-th]].

\bibitem{Kaiser:1995eg}
N.~Kaiser, P.~B.~Siegel and W.~Weise,
%``Chiral dynamics and the low-energy kaon - nucleon interaction,''
Nucl. Phys. A \textbf{594} (1995), 325-345
%doi:10.1016/0375-9474(95)00362-5
%[arXiv:nucl-th/9505043 [nucl-th]].

\bibitem{Oller:1997pn}
J.~A.~Oller and E.~Oset,
%``Chiral symmetry amplitudes in the S wave isoscalar and isovector channels and the sigma, f(0)(980), a(0)(980) scalar mesons,''
AIP Conf. Proc. \textbf{432} (1998) no.1, 824-827
%doi:10.1063/1.55984
%[arXiv:hep-ph/9710554 [hep-ph]].

\bibitem{Pelaez:2021dak}
J.~R.~Pel{\'a}ez, A.~Rodas and J.~Ruiz de Elvira,
%``Precision dispersive approaches versus unitarized chiral perturbation theory for the lightest scalar resonances $\sigma /f_0(500) $ and $\kappa /K_0^*(700) $,''
Eur. Phys. J. ST \textbf{230} (2021) no.6, 1539-1574
%doi:10.1140/epjs/s11734-021-00142-9
%[arXiv:2101.06506 [hep-ph]].



\bibitem{Duan:2020vye} 
M.~Y.~Duan, J.~Y.~Wang, G.~Y.~Wang, E.~Wang and D.~M.~Li, 
%``Role of scalar $a_0(980)$ in the single Cabibbo suppressed process $D^+ \rightarrow \pi ^{+} \pi ^{0} \eta $,'' 
Eur. Phys. J. C \textbf{80} (2020) no.11, 1041 %doi:10.1140/epjc/s10052-020-08630-3 
%[arXiv:2008.10139 [hep-ph]].

\bibitem{Feng:2020jvp} 
X.~C.~Feng, L.~L.~Wei, M.~Y.~Duan, E.~Wang and D.~M.~Li,
%``The a0(980) in the single Cabibbo-suppressed process $\Lambda_c \to \pi^0 \eta p$,'' 
Phys. Lett. B \textbf{846} (2023), 138185 %doi:10.1016/j.physletb.2023.138185 
%[arXiv:2009.08600 [hep-ph]].

\bibitem{Wang:2020pem}
Z.~Wang, Y.~Y.~Wang, E.~Wang, D.~M.~Li and J.~J.~Xie,
%``The scalar $f_0(500)$ and $f_0(980)$ resonances and vector mesons in the single Cabibbo-suppressed decays $\Lambda_c \to p K^+K^-$ and $p\pi^+\pi^-$,''
Eur. Phys. J. C \textbf{80} (2020) no.9, 842
%doi:10.1140/epjc/s10052-020-8347-2
%[arXiv:2004.01438 [hep-ph]].

\bibitem{Black:1999yz}
D.~Black, A.~H.~Fariborz and J.~Schechter,
%``Mechanism for a next-to-lowest lying scalar meson nonet,''
Phys. Rev. D \textbf{61} (2000), 074001
%doi:10.1103/PhysRevD.61.074001
%[arXiv:hep-ph/9907516 [hep-ph]].

\bibitem{LHCb:2019tdw}
R.~Aaij \textit{et al.} [LHCb],
%``Dalitz plot analysis of the $D^+\to K^-K^+K^+$ decay,''
JHEP \textbf{04} (2019), 063
%doi:10.1007/JHEP04(2019)063
%[arXiv:1902.05884 [hep-ex]].

\bibitem{Belle:2020xku}
J.~Y.~Lee \textit{et al.} [Belle],
%``Measurement of branching fractions of  $\Lambda_{c}^{+} \rightarrow \eta\Lambda\pi^{+}$, $\eta \Sigma^{0} \pi^{+}$, $\Lambda(1670) \pi^{+}$, and $\eta \Sigma(1385)^{+}$,''
Phys. Rev. D \textbf{103} (2021) no.5, 052005
%doi:10.1103/PhysRevD.103.052005
%[arXiv:2008.11575 [hep-ex]].

\bibitem{BESIII:2023htx}
M.~Ablikim \textit{et al.} [BESIII],
%``Observation of D+{\textrightarrow}KS0a0(980)+ in the Amplitude Analysis of D+{\textrightarrow}KS0{\ensuremath{\pi}}+{\ensuremath{\eta}},''
Phys. Rev. Lett. \textbf{132} (2024) no.13, 131903
%doi:10.1103/PhysRevLett.132.131903
%[arXiv:2309.05760 [hep-ex]].

\bibitem{BESIII:2024tpv}
M.~Ablikim \textit{et al.} [BESIII], 
%``Observation of D{\textrightarrow}a0(980){\ensuremath{\pi}} in the decays D0{\textrightarrow}{\ensuremath{\pi}}+{\ensuremath{\pi}}-{\ensuremath{\eta}} and D+{\textrightarrow}{\ensuremath{\pi}}+{\ensuremath{\pi}}0{\ensuremath{\eta}},'' 
Phys. Rev. D \textbf{110} (2024) no.11, L111102 %doi:10.1103/PhysRevD.110.L111102
%[arXiv:2404.09219 [hep-ex]].

\bibitem{BESIII:2024mbf}
M.~Ablikim \textit{et al.} [BESIII],
%``Observation of {\ensuremath{\Lambda}}c+{\textrightarrow}{\ensuremath{\Lambda}}a0(980)+ and Evidence for {\ensuremath{\Sigma}}(1380)+ in {\ensuremath{\Lambda}}c+{\textrightarrow}{\ensuremath{\Lambda}}{\ensuremath{\pi}}+{\ensuremath{\eta}},''
Phys. Rev. Lett. \textbf{134} (2025) no.2, 021901
%doi:10.1103/PhysRevLett.134.021901
%[arXiv:2407.12270 [hep-ex]].

\bibitem{BESIII:2025yag} J.~Wu \textit{et al.} [BESIII], %``Observation of an Altered $a_{0}(980)$ Line-shape in $D^{+} \rightarrow \pi^{+}\eta\eta$ due to the Triangle Loop Rescattering Effect,''
[arXiv:2505.12086 [hep-ex]]. 

%\cite{Lyu:2024qgc}
\bibitem{Lyu:2024qgc}
W.~T.~Lyu, S.~C.~Zhang, G.~Y.~Wang, J.~J.~Wu, E.~Wang, L.~S.~Geng and J.~J.~Xie,
%``Evidence of the low-lying baryon {\ensuremath{\Sigma}}*(1/2-) in the process {\ensuremath{\Lambda}}c+{\textrightarrow}{\ensuremath{\eta}}{\ensuremath{\pi}}+{\ensuremath{\Lambda}},''
Phys. Rev. D \textbf{110}, no.5, 054020 (2024)
%doi:10.1103/PhysRevD.110.054020
%[arXiv:2405.09226 [hep-ph]].
%33 citations counted in INSPIRE as of 17 Oct 2026

\bibitem{Lyu:2026ack}
W.~T.~Lyu, S.~W.~Liu, J.~J.~Wu, D.~M.~Li and E.~Wang,
%``Unveiling the elusive $Σ(1380)$ resonance through coupled-channel dynamics in $Λ_c^+\toηπ^+Λ$ reaction,''
[arXiv:2606.04690 [hep-ph]].

%\cite{Lyu:2025oow}
\bibitem{Lyu:2025oow}
W.~T.~Lyu, L.~L.~Wei, D.~M.~Li and E.~Wang,
%``Hadronic decay D+{\textrightarrow}{\ensuremath{\pi}}+{\ensuremath{\eta}}{\ensuremath{\eta}} and the a0(980) and f0(1370) contributions,''
Phys. Rev. D \textbf{112} (2025) no.5, 054015
%doi:10.1103/nfbc-yns2
%[arXiv:2508.04936 [hep-ph]].
%6 citations counted in INSPIRE as of 17 Oct 2026

\bibitem{Zhang:2025lur} 
X.~H.~Zhang, J.~Y.~Zhu, L.~J.~Liu and E.~Wang, %``Roles of a0(980) and a0(1710) in Cabibbo-suppressed process D+{\textrightarrow}{\ensuremath{\pi}}0{\ensuremath{\pi}}+{\ensuremath{\eta}},'' 
Phys. Rev. D \textbf{112} (2025) no.7, 074004 
%doi:10.1103/2nqy-k5lq 
%[arXiv:2508.03493 [hep-ph]].

\bibitem{Ikeno:2024fjr}
N.~Ikeno, J.~M.~Dias, W.~H.~Liang and E.~Oset,
%``$D^+ \rightarrow K_s^0 \pi ^+ \eta $ reaction and $a_0(980)^+$,''
Eur. Phys. J. C \textbf{84} (2024) no.5, 469
%doi:10.1140/epjc/s10052-024-12844-0
%[arXiv:2402.04073 [hep-ph]].

\bibitem{Duan:2024czu}
M.~Y.~Duan, W.~T.~Lyu, C.~W.~Xiao, E.~Wang, J.~J.~Xie, D.~Y.~Chen and E.~Oset,
%``{\ensuremath{\Lambda}}c+{\textrightarrow}{\ensuremath{\eta}}{\ensuremath{\pi}}+{\ensuremath{\Lambda}} reaction and the {\ensuremath{\Lambda}}a0+(980) and {\ensuremath{\pi}}+{\ensuremath{\Lambda}}(1670) contributions,''
Phys. Rev. D \textbf{111} (2025) no.1, 016004
%doi:10.1103/PhysRevD.111.016004
%[arXiv:2410.16078 [hep-ph]].

\bibitem{Xie:2016evi}
J.~J.~Xie and L.~S.~Geng,
%``The $a_0(980)$ and $\Lambda(1670)$ in the $\Lambda^+_c \to \pi^+ \eta$ decay,''
Eur. Phys. J. C \textbf{76} (2016) no.9, 496
%doi:10.1140/epjc/s10052-016-4342-z
%[arXiv:1604.02756 [nucl-th]].

\bibitem{Wang:2022nac}
G.~Y.~Wang, N.~C.~Wei, H.~M.~Yang, E.~Wang, L.~S.~Geng and J.~J.~Xie,
%``Roles of a0(980), {\ensuremath{\Lambda}}(1670), and {\ensuremath{\Sigma}}(1385) in the {\ensuremath{\Lambda}}c+{\textrightarrow}{\ensuremath{\eta}}{\ensuremath{\pi}}+ decay,''
Phys. Rev. D \textbf{106} (2022) no.5, 056001
%doi:10.1103/PhysRevD.106.056001
%[arXiv:2206.01425 [hep-ph]].

\bibitem{Lyu:2024qgc}
W.~T.~Lyu, S.~C.~Zhang, G.~Y.~Wang, J.~J.~Wu, E.~Wang, L.~S.~Geng and J.~J.~Xie,
%``Evidence of the low-lying baryon {\ensuremath{\Sigma}}*(1/2-) in the process {\ensuremath{\Lambda}}c+{\textrightarrow}{\ensuremath{\eta}}{\ensuremath{\pi}}+{\ensuremath{\Lambda}},''
Phys. Rev. D \textbf{110} (2024) no.5, 054020
%doi:10.1103/PhysRevD.110.054020
%[arXiv:2405.09226 [hep-ph]].

%\cite{Geng:2024sgq}
\bibitem{Geng:2024sgq}
C.~Q.~Geng, C.~W.~Liu and S.~L.~Liu,
%``Nonleptonic three-body charmed baryon weak decays with H(15),''
Phys. Rev. D \textbf{109} (2024) no.9, 093002
%doi:10.1103/PhysRevD.109.093002
%[arXiv:2403.06469 [hep-ph]].
%15 citations counted in INSPIRE as of 17 Oct 2026

%\cite{Li:2025gvo}
\bibitem{Li:2025gvo}
M.~Y.~Li, W.~T.~Lyu, L.~J.~Liu and E.~Wang,
%``Roles of the N(1535) and a0(980) in the process {\ensuremath{\Lambda}}c+{\textrightarrow}{\ensuremath{\pi}}+{\ensuremath{\eta}}n,''
Phys. Rev. D \textbf{111} (2025) no.3, 034046
%doi:10.1103/PhysRevD.111.034046
%[arXiv:2501.02859 [hep-ph]].
%11 citations counted in INSPIRE as of 17 Oct 2026

%\cite{BESIII:2026mpt}
\bibitem{BESIII:2026mpt}
M.~Ablikim \textit{et al.} [BESIII],
%``Observation of $ {\Lambda}_c^{+}\to n{\pi}^{+}\eta $ and search for $ {\Lambda}_c^{+}\to n{a}_0{(980)}^{+} $,''
JHEP \textbf{08} (2026), 009
%doi:10.1007/JHEP08(2026)009
%[arXiv:2603.28232 [hep-ex]].
%2 citations counted in INSPIRE as of 09 Oct 2026

%\cite{Miyahara:2016yyh}
\bibitem{Miyahara:2016yyh}
K.~Miyahara, T.~Hyodo, M.~Oka, J.~Nieves and E.~Oset,
%``Theoretical study of the {\ensuremath{\Xi}}(1620) and {\ensuremath{\Xi}}(1690) resonances in {\ensuremath{\Xi}}c{\textrightarrow}{\ensuremath{\pi}}+MB decays,''
Phys. Rev. C \textbf{95} (2017) no.3, 035212
%doi:10.1103/PhysRevC.95.035212
%[arXiv:1609.00895 [nucl-th]].
%65 citations counted in INSPIRE as of 26 Jul 2026

%\cite{Zhang:2026igc}
\bibitem{Zhang:2026igc}
S.~C.~Zhang, W.~T.~Lyu, G.~Y.~Wang, B.~Q.~Ma and E.~Wang,
%``Search for the Low-Lying Excited Baryon {\ensuremath{\Sigma}}$^{*}$(1/2$^{−}$) through Process {\ensuremath{\Lambda}}c+{\textrightarrow}{\ensuremath{\Lambda}}K0{\ensuremath{\pi}}+,''
Chin. Phys. Lett. \textbf{43} (2026) no.5, 050203
%doi:10.1088/0256-307X/43/5/050203
%[arXiv:2601.12778 [hep-ph]].
%2 citations counted in INSPIRE as of 09 Oct 2026

%\cite{Duan:2024okk}
\bibitem{Duan:2024okk}
M.~Y.~Duan, M.~Bayar and E.~Oset,
%``Precise determination of the {\ensuremath{\eta}}{\ensuremath{\Lambda}} scattering length and effective range and relationship to the {\ensuremath{\Lambda}}(1670) resonance,''
Phys. Lett. B \textbf{857} (2024), 139003
%doi:10.1016/j.physletb.2024.139003
%[arXiv:2407.01410 [hep-ph]].
%15 citations counted in INSPIRE as of 09 Oct 2026

%\cite{Gasser:1983yg}
\bibitem{Gasser:1983yg}
J.~Gasser and H.~Leutwyler,
%``Chiral Perturbation Theory to One Loop,''
Annals Phys. \textbf{158} (1984), 142
%doi:10.1016/0003-4916(84)90242-2
%5150 citations counted in INSPIRE as of 17 Oct 2026

%\cite{Scherer:2002tk}
\bibitem{Scherer:2002tk}
S.~Scherer,
%``Introduction to chiral perturbation theory,''
Adv. Nucl. Phys. \textbf{27} (2003), 277
%[arXiv:hep-ph/0210398 [hep-ph]].
%895 citations counted in INSPIRE as of 17 Oct 2026

%\cite{Ren:2015bsa}
\bibitem{Ren:2015bsa}
X.~L.~Ren, E.~Oset, L.~Alvarez-Ruso and M.~J.~Vicente Vacas,
%``Antineutrino induced {\ensuremath{\Lambda}}(1405) production off the proton,''
Phys. Rev. C \textbf{91} (2015) no.4, 045201
%doi:10.1103/PhysRevC.91.045201
%[arXiv:1501.04073 [hep-ph]].
%30 citations counted in INSPIRE as of 17 Oct 2026

%\cite{Xie:2014tma}
\bibitem{Xie:2014tma}
J.~J.~Xie, L.~R.~Dai and E.~Oset,
%``The low lying scalar resonances in the $D^0$ decays into $K^0_s$ and $f_0(500)$, $f_0(980)$, $a_0(980)$,''
Phys. Lett. B \textbf{742} (2015), 363-369
%doi:10.1016/j.physletb.2015.02.006
%[arXiv:1409.0401 [hep-ph]].
%94 citations counted in INSPIRE as of 17 Oct 2026

%\cite{Zhang:2024myn}
\bibitem{Zhang:2024myn}
X.~H.~Zhang, H.~Zhang, B.~C.~Ke, L.~J.~Liu, D.~M.~Li and E.~Wang,
%``Roles of the scalar f0(500) and f0(980) in the process D0{\textrightarrow}{\ensuremath{\pi}}0{\ensuremath{\pi}}0K{\textasciimacron}0,''
Phys. Rev. D \textbf{110} (2024) no.11, 114050
%doi:10.1103/PhysRevD.110.114050
%[arXiv:2409.09966 [hep-ph]].
%8 citations counted in INSPIRE as of 17 Oct 2026


%\cite{Li:2026lbo}
\bibitem{Li:2026lbo}
Y.~Li, E.~Wang, L.~S.~Geng and J.~J.~Xie,
%``Dynamically generated \(N(1535)\) state in the \(\Lambda_{c}^{+}\rightarrow p{\overline{K}}^{0}\pi^{0}\) decay,''
Phys. Rev. D \textbf{113} (2026) no.5, 054039
%doi:10.1103/7tbd-7krm
%[arXiv:2601.13668 [hep-ph]].
%5 citations counted in INSPIRE as of 17 Oct 2026


%\cite{Inoue:2001ip}
\bibitem{Inoue:2001ip}
T.~Inoue, E.~Oset and M.~J.~Vicente Vacas,
%``Chiral unitary approach to S wave meson baryon scattering in the strangeness S = O sector,''
Phys. Rev. C \textbf{65} (2002), 035204
%doi:10.1103/PhysRevC.65.035204
%[arXiv:hep-ph/0110333 [hep-ph]].
%319 citations counted in INSPIRE as of 09 Oct 2026

%\cite{Doring:2005bx}
\bibitem{Doring:2005bx}
M.~Doring, E.~Oset and D.~Strottman,
%``Chiral dynamics in the gamma p ---{\ensuremath{>}} pi0 eta p and gamma p ---{\ensuremath{>}} pi0 K0 Sigma+ reactions,''
Phys. Rev. C \textbf{73} (2006), 045209
%doi:10.1103/PhysRevC.73.045209
%[arXiv:nucl-th/0510015 [nucl-th]].
%102 citations counted in INSPIRE as of 09 Oct 2026

%\cite{Wang:2015pcn}
\bibitem{Wang:2015pcn}
E.~Wang, H.~X.~Chen, L.~S.~Geng, D.~M.~Li and E.~Oset,
%``Hidden-charm pentaquark state in $\Lambda^0_b \to J/\psi p \pi^-$ decay,''
Phys. Rev. D \textbf{93} (2016) no.9, 094001
%doi:10.1103/PhysRevD.93.094001
%[arXiv:1512.01959 [hep-ph]].
%65 citations counted in INSPIRE as of 26 Jul 2026

%\cite{Lu:2016roh}
\bibitem{Lu:2016roh}
J.~X.~Lu, E.~Wang, J.~J.~Xie, L.~S.~Geng and E.~Oset,
%``The $\Lambda_{b}\rightarrow J/\psi K^{0}\Lambda$ reaction and a hidden-charm pentaquark state with strangeness,''
Phys. Rev. D \textbf{93} (2016), 094009
%doi:10.1103/PhysRevD.93.094009
%[arXiv:1601.00075 [hep-ph]].
%94 citations counted in INSPIRE as of 09 Oct 2026

%\cite{Li:2026umb}
\bibitem{Li:2026umb}
M.~Y.~Li, G.~Y.~Wang, N.~C.~Wei, D.~M.~Li and E.~Wang,
%``Probing the isospin structure and low-lying resonances in \(\Lambda_{c}^{+}\rightarrow n{\overline{K}}^{0}\pi^{+}\) decays,''
Phys. Rev. D \textbf{113} (2026) no.9, 094013
%doi:10.1103/s3bp-h492
%[arXiv:2602.12594 [hep-ph]].
%2 citations counted in INSPIRE as of 09 Oct 2026

%\cite{ParticleDataGroup:2024cfk}
\bibitem{ParticleDataGroup:2024cfk}
S.~Navas \textit{et al.} [Particle Data Group],
%``Review of particle physics,''
Phys. Rev. D \textbf{110} (2024) no.3, 030001
%doi:10.1103/PhysRevD.110.030001
%5670 citations counted in INSPIRE as of 26 Jul 2026









\end{thebibliography}

\end{document}